*Spontaneous coordinated activity in cultured networks: Analysis of multiple ignition sites, primary circuits, burst phase delay distributions and functional structures.*

Michael I. Ham[1,2*], Vadas Gintautas[1], Guenter W. Gross[2]

1 Los Alamos National Laboratory, CNLS and T-7, Los Alamos, New Mexico USA
2 Center for Network Neuroscience, University of North Texas, Denton, Texas
* Corresponding author. E-mail address: mih0001@t7.lanl.gov

**Abstract**

All higher order central nervous systems exhibit spontaneous neural activity, though the purpose and mechanistic origin of such activity remains poorly understood. We explore the ignition and spread of collective spontaneous electrophysiological burst activity in networks of cultured cortical neurons growing on microelectrode arrays using information theory and first-spike-in-burst analysis methods. We show the presence of burst leader neurons, which form a mono-synaptically connected primary circuit, and initiate a majority of network bursts. Leader/follower firing delay times form temporally stable positively skewed distributions. Blocking inhibitory synapses usually results in shorter delay times with reduced variance. These distributions are generalized characterizations of internal network dynamics and provide estimates of pair-wise synaptic distances. We show that mutual information between neural nodes is a function of distance, which is maintained under disinhibition. The resulting analysis produces specific quantitative constraints and insights into the activation patterns of collective neuronal activity in self-organized cortical networks, which may prove useful for models emulating spontaneously active systems.

# 1 Introduction

Electrophysiological activity is aways present in neural systems. Such spontaneous activity plays putative roles ranging from synaptic development and maintenance [1,2,3] to anticipatory states [4] which assist animals in reaching rapid decisions with limited sensory input. Understanding the mechanisms of spontaneously generated activity and interaction patterns between neurons are, therefore, issues of substantial importance.

Over several decades, a large body of theoretical analysis and experimental data suggests cortical neuronal networks growing on microelectrode arrays (MEAs) *in vitro* are useful experimental models of neural assembly (e.g. [5,6,7]) though obvious limitations are inherent to extrapolations between *in vitro* and *in vivo* systems [8,9].

In this manuscript we present analysis of neuronal interactions during spontaneous burst activity *in vitro*. These collective high frequency action potential discharges are well documented features of such networks (e.g. [10]) and have been shown to influence learning and information processing by changing synaptic properties [11,12]. Previous research has shown that multiple ignition sites [10,13,14] recruit network neurons to create network bursts. Here, temporal relationships between leader (first neuron to fire in a network burst) and follower neurons are examined using a first-spike-in-burst analysis method to create response delay distributions (RDDs). Disinhibition with bicuculline reveales changes in ignition site statistics and follower responsiveness.

RDD features are found to provide estimates of the distance between leader and follower neurons during network activation. Similarly, mutual information between neuronal pairs is shown to be correlated with the distance between the two neurons in each pair.

Our approach reveals new insight into functional connectivity and network organization which may be useful for creating models of small to medium sized neural networks.

# 2 Methods

## 2.1 Microelectrode array fabrication

In-house MEA fabrication is described in previous publications [15]). Briefly, glass plates with a 100-nanometer layer of indium-tin-oxide (ITO, Applied Films Corp., Boulder, CO) were photo etched to create a recording matrix of 64 electrodes measuring 8-10 um in width and conductors leading to peripheral amplifier contacts on a 5 x 5 cm glass plate. Plates were spin-insulated with methyltrimethoxysilane resin, cured, and de-insulated at the electrode tips with single laser shots. Exposed electrode terminals were electroplated with colloidal gold to decrease impedance at 1 kHz to approximately 0.8 MΩ. Butane flaming followed by application of poly-D-lysine and laminin helped cell adhesion. These microelectrode arrays, featuring substrate integrated thin film conductors allow long-term, extracellular microvolt recording of action potentials from 64 discrete sites in a neuronal network.

## 2.2 Cell culture

Frontal cortices were dissected from 16 to 17 day old mouse embryos. The tissue was mechanically minced, enzymatically dissociated, triturated, and combined with medium (Dulbecco's Modified Eagles Medium (DMEM) supplemented with 10% fetal bovine serum and 10% horse serum). Dissociated cells (100k / ml) were seeded on MEAs with medium addition after cells had adhered (usually 2-3 hrs). After 5 days, cultures were fed DMEM supplemented with 5% horse serum (DMEM-5). Greater detail is provided in earlier publications [16]. Cultures were incubated at 37 °C in a 10% CO2 atmosphere with 50% medium changes performed twice a week until used for experiments.

## 2.3 Electrophysiological Data Acquisition

For electrophysiological recordings, cultures were assembled into a recording apparatus on an inverted microscope connected to a two-stage, 64 channel amplification and signal processing system (Plexon Inc., Dallas). Cultures were maintained at a temperature of 37 °C, and pH of 7.4. The pH was maintained by a 10 ml / min flow of 10% CO2 in air into a cap on the chamber block featuring a heated ITO window to prevent condensation. Water evaporation was compensated by a syringe pump (Harvard Instruments) with the addition of 60 to 70 uL / hr sterile water. Details of chamber assembly and recording procedures can be found in previous papers [16]. Total system gain was set at 10k and action potential (AP) activity with a sampling resolution of 25 µs was recorded for later analysis.

## 2.4 Electrophysiological Data Acquisition

To identify network bursts, recordings are partitioned into 10 ms bins and number of spikes per bin is determined. An upper and lower threshold algorithm is used to identify network bursts. Upper threshold is selected as 20% of all recorded spikes. The lower threshold is found by rounding up the average bin count and is generally set at 1 or 2 spikes/bin.

The algorithm finds the first bin count at least equal to the lower threshold. Then, two possible scenarios are examined: (1) the number of spikes is greater than or equal to the upper threshold which indicates the start of a global burst. Bursts continue while consecutive bin counts are at or above the lower threshold. (2) Upper threshold is not reached. Bin is marked as the potential beginning of a burst. Consecutive 10 ms bins are searching for one that satisfies the upper threshold (a burst), or falls below the lower threshold (no burst).

All network bursts end when a consecutive bin falls below the lower threshold. Network bursts with activity gaps of less than 100 ms are combined. The first neuron in each burst (burst leader) to fire is recorded. The first spike each follower made in each network burst is used to determine response (phase) delays between leaders and followers.

## 2.5 Informational relationships

To quantify informational relaionships, neuronal time series are digitized into 10 ms bins. If a neuron fired within a bin, the bin is assigned a 1. If it did not fire, the bin is assigned a 0. Mutual information (MI) between neurons X and Y is calculated using $MI=\Sigma p(X,Y)*\log2[p(X,Y)/p(X)p(Y)]$ where p(X,Y) is the joint probability distribution and p(X) and p(Y) are the single variable marginals [17,18]. Normalized nutual information is found by dividing the MI by the smaller of the Shannon entropy of X or Y. Mutual information is 0 if and only if neurons X and Y are independent of each other.

# 3 Results

**Major burst leaders**

In 10 experimental networks, each recorded for at least three hours, all neurons led at least one network burst. However, only a small percentage, ( average 17%) are found to be major burst leaders (MBLs, Fig 1). Major burst leaders are defined as neurons that led at least 4% (arbitrarily chosen) of all network bursts. The set of MBLs are found to be relatively stable over many hours, though individual leadership rates fluctuated [19]. On average, MBLs led 84% of all network bursts (Fig 1).

Neuronal spike rates do not appear directly linked to burst leadership. In Fig 1, the percent spike activity (percent of total activity a neuron contributes) is compared to burst leadership under both native and bicuculline activity. It should be noted that MBLs are some of the most active recorded neurons, but that activity is not a good predictor of leadership.

**Response delay distributions**

For each MBL-follower pair, the aggregate of their response delays (time between MBL starting a burst, and follower responding) is represented as a response delay distribution (RDD, Fig2). RDDs are observed to be highly variable in nature and unique for each leader/follower pair. Three important features are extracted from a pair's RDD. (1) Minimum response delay (MRD) values. These are identified as the intersection of the rising edge of the distribution with a value equal to 10% of the distributions peak. MRDs represent the minimum time it takes a signal to travel from a leader to a follower. Most MRDs are approximately 2 ms, but ranged up to 20 ms. (2) Peak delay. The most probable delay time between leader and follower onset. (3) Paired response correlation (PRC). A measure of how likely a follower is to respond to a given leader.

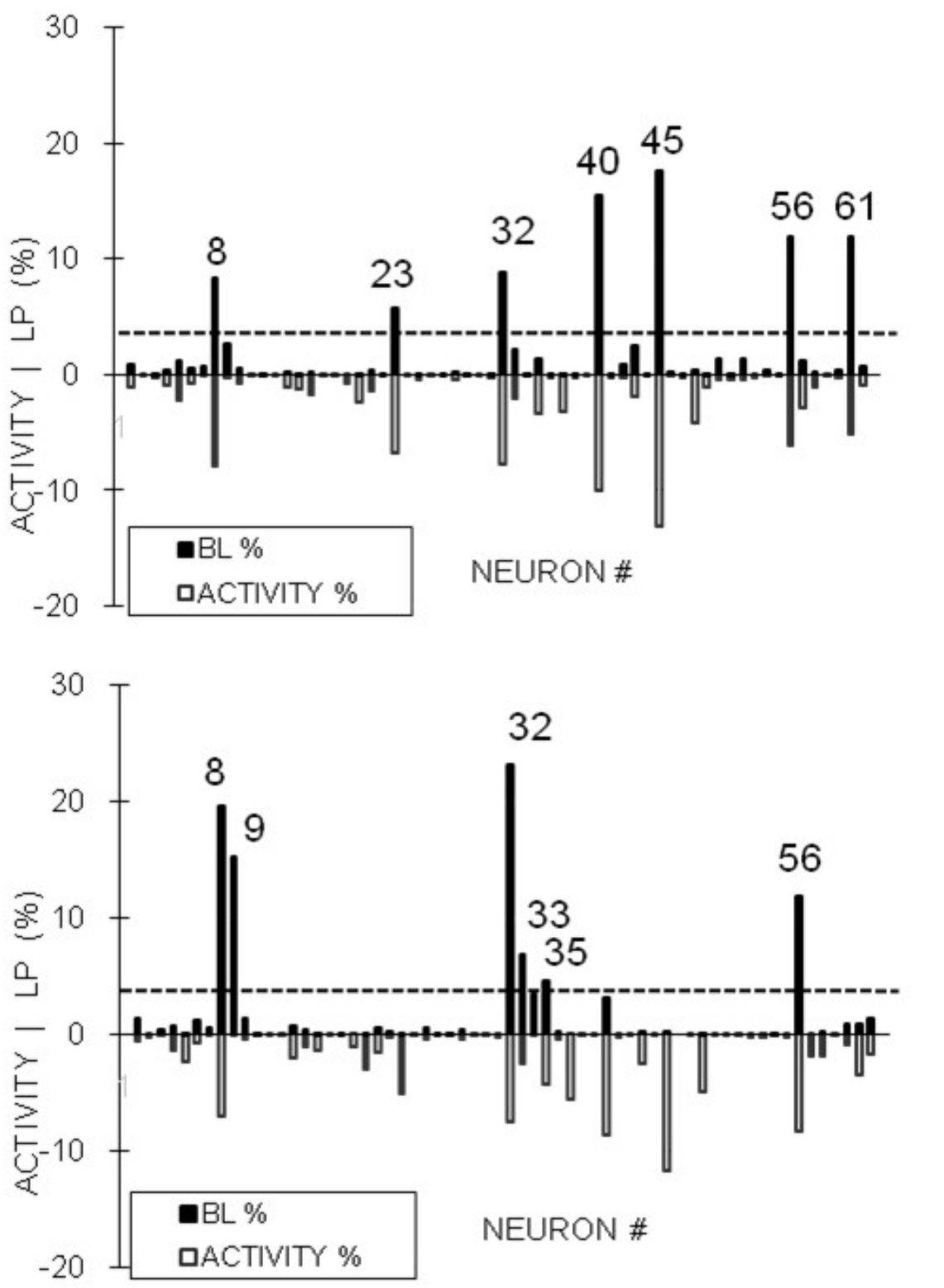

**Fig. 1.** Burst leadership (positive) and activity probability (negative) is shown for neurons recorded in native activity (top) and under 40 µM bicuculline (bottom). Neurons leading with rates above an arbitrarily chosen cutoff (dashed line) are considered major burst leaders. The addition of bicuculline changed the pool of major burst leaders which are stable during native activity.

**Primary circuit**

All MBLs exhibit high PRC (>70%) values with respect to all other MBLs. Additionally MBLs have a PRC value > 90% with respect to at least one other MBL. Few non MBLs have PRC values of this magnitude. RDDs of MBL pairs show MRDs around 2 ms, which suggests that the shortest path between two MBLs is a single synapse [20]. Therefore, we conclude that MBLs form a highly connected 'primary circuit' responsible for initiating the majority of all network bursts and maintaining long term spontaneous activity.

**Disinhibition with bicuculline**

Disinhibition with 40µM bicuculline changes the composition of MBL pools (Fig 1). In eight networks where bicuculline was added, there are 52 MBLs during native activity and 48 after. Only 25 of the latter are MBLs during both time periods. Bicuculline also changes the nature of the response delay distributions. Four unique changes are observed. (1) increased responsiveness to the MBL, (2) overall shift of the distribution to shorter phase delays, (3) response by previously unresponsive neurons, (4) rarely, peak and MRD shift to higher phase delays. There are no observed cases in which a follower became less responsive to a burst leader. All four RDD changes could happen to followers of the same burst leader.

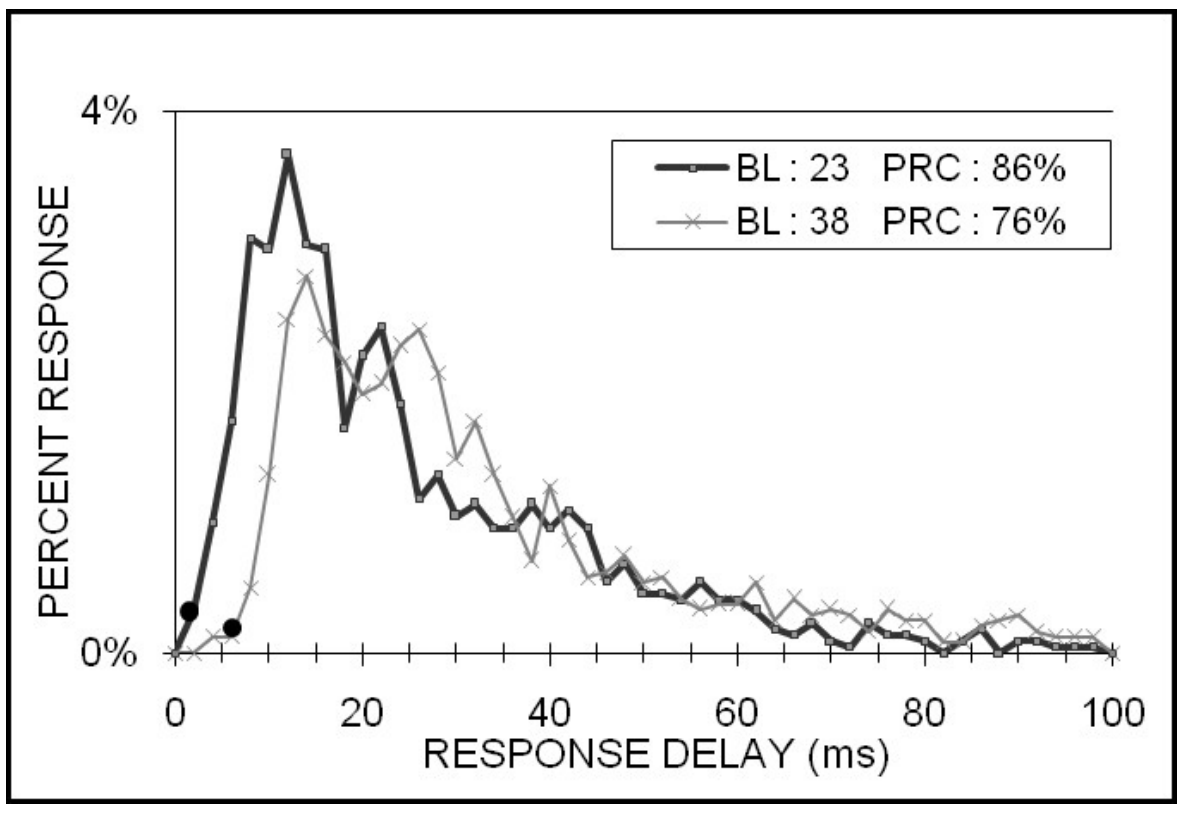

**Fig. 2.** Response delay distribution for a single follower and two burst leaders (2ms bins). Minimum response delays are depicted with filled circles. Followers have unique responses to different burst leaders.

**Mutual information and distance**

Previously, we showed that RDD features are affected by distance from the burst leader [19]. Here we apply an information theoretic approach to examine whether all neuronal relationships are impacted by distance from one another. In Fig 3, the mutual information for all neural pairs is calculated first under native conditions, then after application of 40 µM bicuculline. Values are then averaged for all exact distances. The mean is plotted as a point and the standard error is indicated with error bars. We observe that, on average, mutual information between neuronal pairs drops as distance increases.

# 4 Discussion

We show that major burst leaders play an important role in network activation. Together, they dominate the initiation of spontaneous network firing pattens by exciting the network. However activity from a single MBL may not be enough to trigger the rest of the network. Our finding that MBLs are well connected to other MBLs through the primary circuit, suggests that leadership may be a shared property, requiring the combined activity of several highly connected neurons.

Response delay distributions (RDDs) between leader/follower pairs reveal the nature of the relationship between these two neurons. Small minimum response delays correspond to the presence of short synaptic pathways between a pair. Peak delay times show the most probable path delay. Paired response correlation show how effective a particular leader is at activating a follower. These values can be very different for leaders of the same follower, suggesting that recruitment may depend heavily on which neruon leads a network burst.

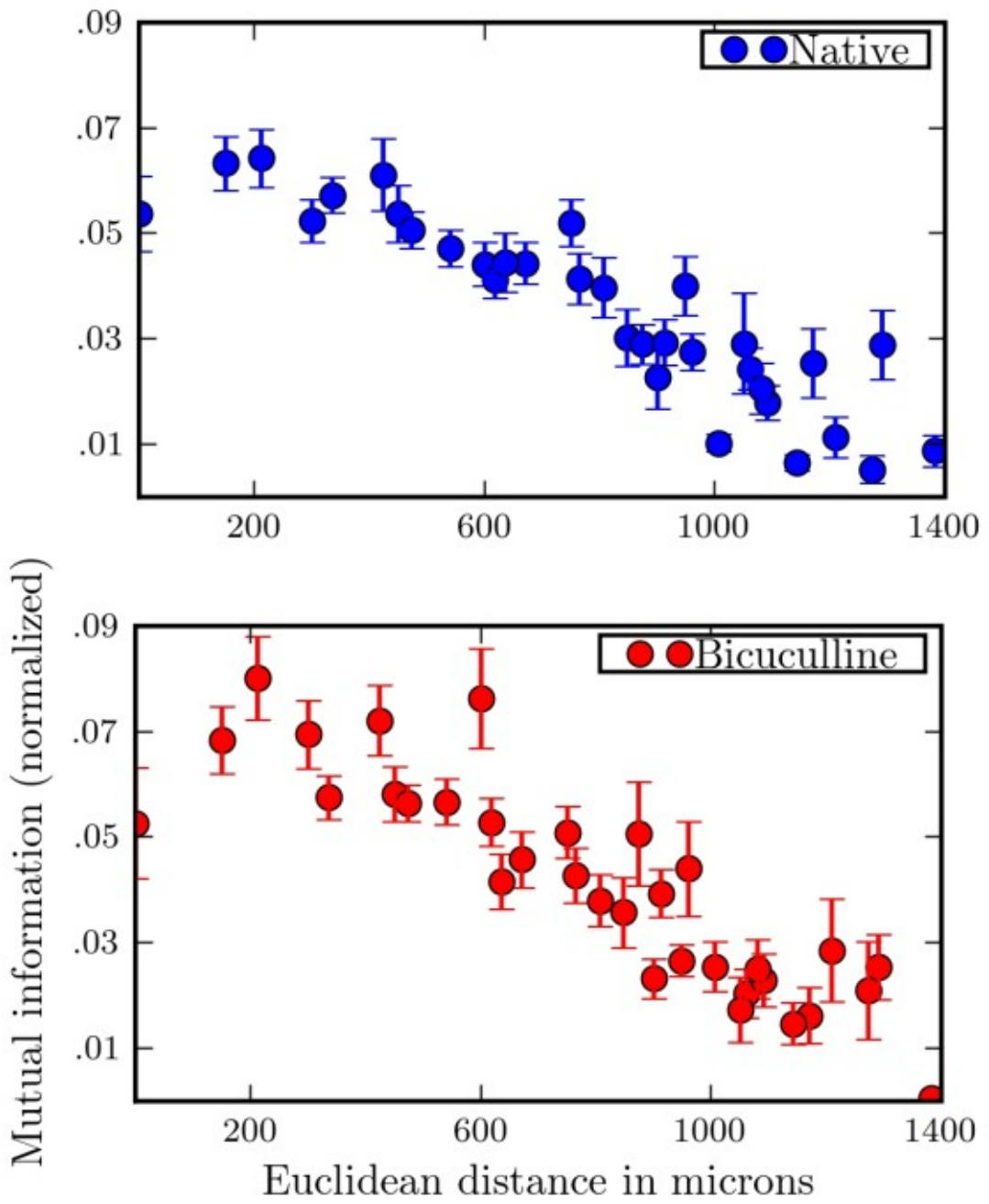

**Fig. 3.** Average mutual information values and the distance between neuronal pairs. The mutual information is measured before (top) and after (bottom) the addition of 40 μM bicuculline. Information decreases as a function of distance in both cases.

Blocking inhibition with bicuculline led to an immeadiate change in the burst leadership pool and response delay distributions. The abruptness of this change suggests that existing network circuitry remains unchanged while only computational activities are modified.

In a previous paper, we showed that RDD characteristics are a function of distance [19]. Here we show that mutal information between all neurons is also a function of distance. While not suprising, these two results show that burst propogation and mutual information are both proportional to distance.

Network bursts are well known features of natural nervous system activity. We hope that characterizations of collective activity patterns in these cultures will guide new models and lead to improved biologically accurate models of the computational abilities of the nervous system.

**Acknowledgement**

This project supported by LDRD-ER-20050411, and in part by the Texas Advanced Technology Program. The authors also wish to thank Luis Bettencourt for providing invaluable insight and expertise.

**References**

[1] R.R.Provine (1979) Wing-flapping develops in wingless chicks. *Behav Neur Biol, 27:233–237.*

[2] M. Meister, R.O.L. Wong, D.A. Baylor, and C.J. Shatz (1991) Synchronous bursts of action potentials in ganglion cells of the developing mammalian retina. *Science, 252:939–943.*

[3] N.C. Spitzer (2006) Electrical activity in early neuronal development. *Nature, 444:707–712*

[4] Y. Takikawa, R. Kawagoe, and O. Hikosaka (2002) Reward-dependent spatial selectivity of anticipatory activity in monkey caudate neurons. *J Neurophysiol, 87:508–515.*

[5] T. B. DeMarse, D. A. Wagenaar, A. W. Blau, and S. M. Potter (2001) The neurally controlled animat: Biological brains acting with simulated bodies. *Auton. Rob., 11:305*

[6] G. Shahaf and S. Marom (2001) Learning in networks of cortical neurons. *J Neurosci, 21:8782–8788*

[7] L. M. A. Bettencourt, G. J. Stephens, M. I. Ham, and G. W. Gross (2007) Functional structure of cortical neuronal networks grown in vitro. *Phys. Rev. E, 75(2):021915*

[8] M.A. Corner, J. Van Pelt, P.S. Wolters, R.E. Baker, and R.H. Nuytinck (2002) Physiological effects of sustained block-ade of excitatory synaptic transmission on spontaneously active developing neuronal networks?an inquiry into the reciprocal linkage between intrinsic biorhythms and neuroplasticity in early ontogeny. *Neurosci Biobehav Rev, 26:127–185*

[9] S. Marom and G. Shahaf (2002) Development, learning and memory in large random networks of cortical neurons: lessons beyond anatomy. *Q. Rev. Biophys., 35:63*

[10] E. Maeda, H. P. C. Robinson, and A. Kawana (1995) The mechanisms of generation and propogation of synchronized bursting in developing networks of cortical neurons. *J. Neurosci, 15:939*

[11] T.V. Bliss and G.L.A Collingridge (1993). A synaptic model of memory: long-term potentiation in the hippocampus. *Nature, 361:31–39*

[12] A. Artola and W. Singer (1994) NMDA receptors and developmental plasticity in visual neocortex. In: The NMDA receptor. *London: Oxford UP*

[13] D. Eytan and S. Marom. (2006) Dynamics and effective topology underlying synchronization in networks of cortical neurons. *J Neurosci, 26:8465–8476*

[14] O. Feinerman, M. Segal, and E. Moses (2007) Identification and dynamics of spontaneous burst initiation zones in unidimensional neuronal cultures. *J Neurophysiol, 97(4):2937–2948*

[15] G.W. Gross, W.Y. Wen, and J.W. Lin (1985) Transparent indium-tin oxide electrode patterns for extracellular, multisite recording in neuronal cultures. *Neurosci Methods, 15(3):243–252*

[16] G.W. Gross (1994) Internal dynamics of randomized mammalianneuronal networks in culture. In: Enabling technologies for cultured neural networks. *New York: Academic*

[17] T. M. Cover, and J. A.. Thomas, Elements of Information Theory *(Wiley, New York, 1991)*

[18] L.M.A. Bettencourt, V. Gintautas and M.I. Ham (2008) Identification of functional information subgraphs in complex networks. *to apear in Phys. Rev. Lett.*

[19] M.I. Ham, L.M.A. Bettencourt, F.D. McDaniels, G.W. Gross (2008) Spontaneous coordinated activity in cultured networks: analysis of multiple ignition sites, primary circuits, and burst phase delay distributions. *J. Comp Neurosci 24(3) 346-357*

[20] K. Nakanishi and F. Kukita (1998). Functional synapses in-synchronized bursting of neocortical neurons in culture. *Brain Res, 795:137–146.*